\documentclass[12pt]{article}

\newcommand\lsim{\mathrel{\rlap{\lower4pt\hbox{\hskip1pt$\sim$}}
    \raise1pt\hbox{$<$}}}
\newcommand\gsim{\mathrel{\rlap{\lower4pt\hbox{\hskip1pt$\sim$}}
    \raise1pt\hbox{$>$}}}

\usepackage{graphics}
\usepackage{epsfig}
\usepackage{psfrag}

\begin{document}

\sf
\centerline{\Huge Probing seesaw at LHC}

\vspace{7mm}

\centerline{\large
Borut Bajc$^{1}$, 
Miha Nemev\v sek$^{1}$ and 
Goran Senjanovi\'c$^{2}$}
\vspace{1mm}
\centerline{$^{1}$ {\it\small J.\ Stefan Institute, 1001 Ljubljana, Slovenia}}
\centerline{$^{2}${\it\small International Centre for Theoretical Physics,
Trieste, Italy}}

\vspace{5mm}
   
\centerline{\large\sc Abstract}
\begin{quote}
\small

We have recently proposed a simple SU(5) theory with an adjoint 
fer\-mio\-nic multiplet on top of the usual minimal spectrum. This 
leads to the hybrid scenario of both type I and type III seesaw 
and it predicts the existence of the fermionic SU(2) triplet 
between $100$ GeV and $1$ TeV for a conventional GUT scale of about 
$10^{16}$ GeV, with main decays into $W$ ($Z$) and leptons, 
correlated through Dirac Yukawa couplings, and lifetimes 
shorter than about $10^{-12}$ sec. These decays are lepton number violating 
and they offer an exciting signature of $\Delta L=2$ dilepton events 
together with 4 jets at future $pp$ ($p\bar p$) colliders. 
Increasing the triplet mass 
endangers the proton stability and so the seesaw mechanism could 
be directly testable at LHC.


\end{quote}
\rm

\section{Introduction}

The seesaw mechanism \cite{seesaw} has been recognized as the 
most natural scenario for understanding the smallness of neutrino mass. 
It implies the existence of heavy particles, which after being 
integrated out, lead to the gauge invariant operator \cite{Weinberg:1979sa}

\begin{equation}
\label{weinberg}
{\cal L}_{eff}=y_{eff}\frac{LLHH}{M}\;,
\end{equation}

\noindent 
with $M\gg M_W$ usually assumed. As shown in \cite{Ma:1998dn}, 
there are three different types 
of heavy particles that can induce (\ref{weinberg}):

I) SM fermionic singlets, coupled to leptons through Dirac Yukawa 
couplings and usually called right-handed neutrinos (type I seesaw) 
\cite{seesaw};

II) SU(2) bosonic triplet ($Y=2$) coupled to leptons through 
Majorana type couplings (type II seesaw) \cite{Magg:1980ut};

III) SU(2) fermionic triplet ($Y=0$) coupled to leptons through 
Dirac Yukawas, just like the singlet ones in I) (type III seesaw) 
\cite{Foot:1988aq}.

Whatever one chooses, one needs a predictive theory above the SM in 
order to shed some light on neutrino masses; otherwise, one can as well 
stick to the effective operator in (\ref{weinberg}). The best bet for 
such a theory is grand unification since it can predict new mass 
scale(s). It turns out that both type I and type II seesaw find their 
natural role in SO(10) theory due to the automatically present 
left-right symmetry 
\cite{Pati:1974yy,Mohapatra:1974gc,Senjanovic:1975rk,Senjanovic:1978ev}. 
Although SO(10) is sufficient by 
itself to determine all the parameters in the I) and II) cases , 
and even  the 1-3 mixing angle \cite{so10rev}, 
the check of the seesaw is only indirect: one can  at best 
relate neutrino properties to proton decay. The main point is that 
both right-handed neutrinos and the SU(2) scalar triplet are predicted 
to be very heavy, close to the GUT scale. 

What about the type III seesaw? It is clearly custom fit for the SU(5) 
theory, as suggested recently \cite {Bajc:2006ia},
since it only requires adding the adjoint fermions $24_F$ to the 
existing minimal model with three generations of quarks and leptons, 
and $24_H$ and $5_H$ Higgs fields. This automatically leads to the hybrid 
scenario of both type I and type III seesaw, since $24_F$ has also a SM 
singlet fermion, i.e. the right-handed neutrino.  One ends up with 
 a realistic spectrum of two massive and one massless light neutrino. 
The massless one can of course pick up a tiny mass due to say Planck scale
effects \cite{Akhmedov:1992hh} or running effects \cite{Davidson:2006tg}, 
too small to play any direct phenomenological role.

The main prediction of this theory is the lightness of the fermionic 
triplet (for a recent alternative scenario with light triplets see 
\cite{Gudnason:2006mk}). For a conventional value of 
$M_{GUT}\approx 10^{16}$ GeV, the 
unification constraints strongly suggest its mass below TeV, relevant 
for the future colliders such as LHC. The triplet fermion 
decay predominantly into $W$ (or $Z$) and leptons, with lifetimes shorter 
that about $10^{-12}$ sec.

Equally important, the decays of the triplet are dictated by the same 
Yukawa couplings that lead to neutrino masses and thus one has an 
example of predicted low-energy seesaw directly testable at colliders 
and likely already at LHC.

In this expanded version of the original work we sistematically study 
the spectrum and the couplings of the theory. In the next section we 
focus on the unification constraints on the particle spectrum. 
We perform a numerical study using two-loop RGE taking into account 
various mass scales of the theory. 
We discuss $b-\tau$ unification and the predictions of the fermionic 
triplet mass depending on the GUT scale. We find a maximal value of 
the GUT scale: $M_{GUT}\le 10^{16}$ GeV, which offers a great hope of 
observing proton decay in a not so distant future. The color octets 
turn out not to be light enough for direct observation.

In section III we focus on the phenomenological implications of the 
theory for LHC. We discuss carefully the decay modes of the triplets 
and their connection with neutrino masses and mixings. Whereas for 
generic values of Yukawa couplings it is not easy to make clear 
predictions, for the case of vanishing $\theta_{13}$ or large Yukawa 
couplings (possibly related to large flavour violating processes) 
one can constraint the relevant branching ratios and thus 
directly test the seesaw mechanism at colliders.

Next, in section IV we turn our attention to cosmology and discuss 
leptogenesis. We find that it can work only in the resonant regime 
which implies the same mass of the fermionic triplet and singlet 
and further constrains the parameters space of the theory. The 
nice feature of a high degree of predictivity of this theory has also 
a negative implication: we  show that there is no stable particle 
candidate for the dark matter of the universe. We conclude our work 
with section V, where we also discuss the relevance of our work for 
supersymmetry.

\section{Unification constraints and the mass scales of the theory}

The minimal 
implementation of the type III seesaw in 
nonsupersymmetric SU(5) requires a fermionic adjoint 
$24_F$ in addition to the usual field content $24_H$, $5_H$ 
and three generations of fermionic $10_F$ and $\overline{5}_F$. 
The consistency of the charged fermion masses requires 
higher dimensional operators in the usual Yukawa sector 
\cite{Ellis:1979fg}. One must add the new Yukawa interactions 

\begin{eqnarray}
\label{lyukawanu}
{\cal L}_{Y\nu}&=&
y_0^i\bar 5_F^i24_F5_H\\
&+&\frac{1}{\Lambda}\bar 5_F^i
\left(
y_1^i24_F24_H
+y_2^i24_H24_F+
y_3^iTr 24_F24_H
\right)
5_H+h.c.\;.\nonumber
\end{eqnarray}

\noindent
After the SU(5) breaking 

\begin{equation}
\langle 24_H\rangle =\frac{M_{GUT}}{\sqrt{30}}
diag\left(2,2,2,-3,-3\right)
\end{equation}

\noindent
one obtains the following physical 
relevant Yukawa interactions for neutrino with the triplet 
$\sigma_3^F\equiv\overrightarrow{\sigma}_3^F\overrightarrow{\tau}$ 
(type III) and singlet $\sigma_0^F$ (type I) fermions:

\begin{equation}
\label{lynu}
{\cal L}_{Y\nu}=L_i\left(
y_T^i\sigma_3^F+
y_S^i\sigma_0^F\right)H+h.c.\;,
\end{equation}

\noindent
where $y_T^i$, $y_S^i$ are two different 
linear combinations of $y_0^i$ and $y_a^i M_{GUT}/\Lambda$ 
($a =1,2,3$). 
It is clear from the above formula that besides the new 
appearence of the triplet fermion, the singlet fermion 
in $24_F$ acts precisely as the right-handed neutrino; 
it should not come out as a surprise, as it has the right 
SM quantum numbers. 

Even before we discuss the physical consequences in detail, 
one important prediction emerges: only two light neutrinos 
get mass, while the third one  remains massless. 

In order to discuss the masses of the new fermions, we need 
the new Yukawa couplings between $24_F$ and $24_H$

\begin{eqnarray}
\label{lf}
{\cal L}_{F}&=&m_FTr 24_F^2+
\lambda_FTr 24_F^224_H\\
&+&\frac{1}{\Lambda}\left(
a_1Tr 24_F^2 Tr 24_H^2 
+a_2\left(Tr 24_F24_H \right)^2
\right.\nonumber\\
&+&\left.
a_3Tr 24_F^224_H^2 +
a_4Tr 24_F24_H24_F24_H \right)\;,\nonumber
\end{eqnarray}

\noindent
where we include the higher dimensional terms for the sake 
of consistency. The masses of the new fermions are 

\begin{eqnarray}
\label{m0}
m_0^F&=&m_F-\frac{\lambda_FM_{GUT}}{\sqrt{30}}+
\frac{M_{GUT}^2}{\Lambda}\left[a_1+a_2+
\frac{7}{30}\left(a_3+a_4\right)\right]\;,\\
\label{m3}
m_3^F&=&m_F-\frac{3\lambda_FM_{GUT}}{\sqrt{30}}+
\frac{M_{GUT}^2}{\Lambda}\left[a_1+
\frac{3}{10}\left(a_3+a_4\right)\right]\;,\\
\label{m8}
m_8^F&=&m_F+\frac{2\lambda_FM_{GUT}}{\sqrt{30}}+
\frac{M_{GUT}^2}{\Lambda}\left[a_1+
\frac{2}{15}\left(a_3+a_4\right)\right]\;,\\
\label{m32}
m_{(3,2)}^F&=&m_F-\frac{\lambda_FM_{GUT}}{2\sqrt{30}}+
\frac{M_{GUT}^2}{\Lambda}\left[a_1+
\frac{\left(13a_3-12a_4\right)}{60}\right]\;.
\end{eqnarray}

Next we turn to the bosonic sector of the theory. We will 
need the potential for the heavy field $24_H$

\begin{equation}
\label{v24h}
V_{24_H}=
m_{24}^2Tr 24_H^2 +
\mu_{24}Tr 24_H^3 
+
\lambda_{24}^{(1)}Tr 24_H^4 +
\lambda_{24}^{(2)}\left(Tr 24_H^2 \right)^2
\;,
\end{equation}

\noindent
and its interaction with the light fields

\begin{eqnarray}
\label{v5h}
V_{5_H}&=&m_H^25_H^\dagger 5_H+
\lambda_H \left(5_H^\dagger 5_H\right)^2
+\mu_H5_H^\dagger 24_H5_H\nonumber\\
&+&
\alpha 5_H^\dagger 5_H Tr 24_H^2 +
\beta5_H^\dagger 24_H^25_H\;.
\end{eqnarray}

It is a straightforward exercise to show that the 
masses of the bosonic triplet and octet are arbitrary and 
that one can perform the doublet-triplet splitting through 
the usual fine-tuning. However splitting its mass from the triplet and 
the octet fermion masses require the inclusion of higher 
dimensional terms, which in turn gives an upper bound to the 
mass of the leptoquark 

\begin{equation}
\label{m32lambda}
m_{(3,2)}^F\lsim\frac{M_{GUT}^2}{\Lambda}\;,
\end{equation}

\noindent
where $\Lambda$ is the cutoff of the theory. One could take naively 
$\Lambda$ on the order of the Planck scale, since the theory is 
asymptotically free. However, without higher dimensional operators 
one predicts $m_b = m_{\tau}$ at the GUT scale \cite{Chanowitz:1977ye}, 
which fails badly, as much as in the standard model, and thus one 
must take a lower cut-off \footnote{We thank Ilja Dor\v sner for 
pointing it to us. For further details see \cite{Dorsner:2006fx}.}. 
To see this we did a one-loop 
Yukawa running, with a two-loop gauge running. The result 
$y_\tau\approx 0.01$ and the ratio $y_b/y_\tau\lsim 0.65$ 
is valid for any physically allowed value of $M_{GUT}$. 

Thus the analysis requires a cut-off at most two 
orders of magnitude above the GUT scale. In what follows we 
take $\Lambda=100\;M_{GUT}$ to ensure the correct mass 
relations and maximize perturbativity (for a lower cutoff 
see \cite{Dorsner:2006fx}).

We are now fully armed to study the constraints on 
the particle spectrum by performing the renormalization 
group analysis. For the sake of illustration we present 
first the one-loop analysis. From \cite{Bajc:2006ia} one has 

\begin{eqnarray}
\label{a12}
&&\exp{\left[30\pi\left(\alpha_1^{-1}-\alpha_2^{-1}\right)
\left(M_Z\right)\right]}=\\
&&\left(\frac{M_{GUT}}{M_Z}\right)^{84}
\left(\frac{\left(m_3^F\right)^4
m_3^B}{M_Z^5}\right)^{5}
\left(\frac{M_{GUT}}{m_{(3,2)}^F}\right)^{20}
\left(\frac{M_{GUT}}{m_T}\right)\;,\nonumber\\
\label{a13}
&&\exp{\left[20\pi\left(\alpha_1^{-1}-\alpha_3^{-1}\right)
\left(M_Z\right)\right]}=\\
&&\left(\frac{M_{GUT}}{M_Z}\right)^{86}
\left(\frac{\left(m_8^F\right)^4
m_8^B}{M_Z^5}\right)^{5}
\left(\frac{M_{GUT}}{m_{(3,2)}^F}\right)^{20}
\left(\frac{M_{GUT}}{m_T}\right)^{-1}\;,\nonumber
\end{eqnarray}

\noindent
where $m_3^{F,B}$, $m_8^{F,B}$, $m_{(3,2)}^F$ and $m_T$ are 
the masses of weak triplets, colour octets, (only fermionic) 
leptoquarks and (only bosonic) colour triplets respectively. 

From the well known problem in the standard 
model of the low meeting scale of $\alpha_1$ and $\alpha_2$, 
it is clear that the SU(2) triplet should be as light as 
possible and the colour triplet as heavy as possible. In 
order to illustrate the point, take $m_3^F=m_3^B=M_Z$ and 
$m_T=M_{GUT}$. This gives ($\alpha_1^{-1}(M_Z)=59$,
$\alpha_2^{-1}(M_Z)=29.57$, $\alpha_3^{-1}(M_Z)=8.55$) 
$M_{GUT}\approx 10^{15.5}$ GeV. Increasing the 
triplet masses $m_3^{F,B}$ reduces $M_{GUT}$ dangerously, making 
 proton decay too fast. 

For more reliable results one needs a two-loop analysis. 
We focused on the following regions in parameter space:

1) $m_8^{F,B}> 10^5$ GeV to comply with cosmological bounds 
coming from nucleosynthesis. This limit is analogous to the 
limit on the sfermion masses in split supersymmetry 
\cite{Arkani-Hamed:2004fb,Giudice:2004tc} coming 
from gluino lifetime \cite{Arvanitaki:2005fa}. At the time 
of nucleosynthesis all colour octets should have already decayed 
into a righthanded quark and an off-shell colour triplet 
through the Yukawa interactions (\ref{lyukawanu});

2) $m_T> 10^{12}$ GeV from proton decay;

3) $M_{GUT}> 10^{15.5}$ GeV again from proton decay; 

4) $m_{(3,2)}^F< M_{GUT}^2/\Lambda=M_{GUT}/100$ from (\ref{m32lambda}) 
and the above discussion on the choice of the cutoff.

The two-loop analysis maintain an approximate dependence on 
the combinations $m_3\equiv\left((m_3^F)^4m_3^B\right)^{1/5}$ 
and  $m_8\equiv\left((m_8^F)^4m_8^B\right)^{1/5}$ 
as at 1-loop order (\ref{a12}), (\ref{a13}). This is useful 
in the numerical analysis, since one can first use as 
varying parameters just these combinations, and the extrapolate 
the result for the case of different fermionic and bosonic masses. 

\begin{figure}[!ht]
  \centering
  \psfrag{x}{$\log_{10} \left(\frac{M_{GUT}}{\rm GeV}\right)$}
  \psfrag{y}{$\log_{10} \left(\frac{m_3^{max}}{\rm GeV}\right)$}
  \includegraphics[totalheight=5.5cm]{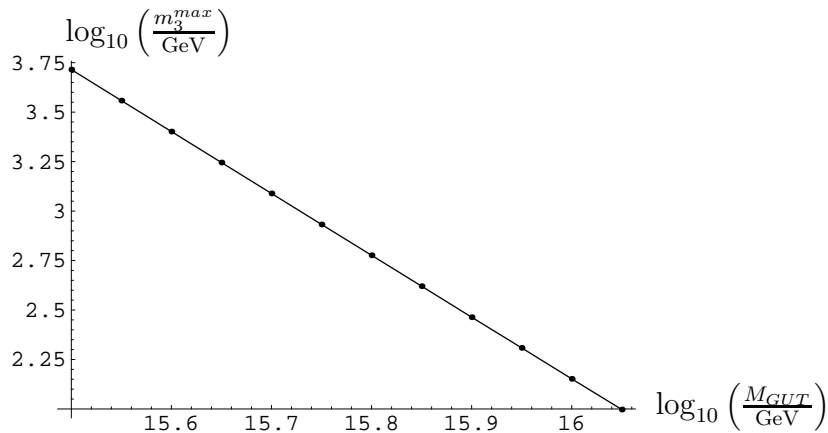}
\caption{\label{fig1} The maximum value of the effective triplet 
mass $m_3$ as a function of the unification scale $M_{GUT}$ from 
the two-loop analysis.}
\end{figure}

We have seen that at 1-loop order the mass of the fermionic triplet 
is predicted to lie below TeV. This bound gets somewhat relaxed at 
2-loop order, as can be seen from Fig. \ref{fig1}. 

The fermionic triplet can be even higher at the price of lowering 
the bosonic triplet. It must be stressed although, that these maximal 
values are not typical: one must stretch the parameters, i.e. go to 
some corner in parameter space to evade the 1-loop bounds. In 
other words, in most of the parameter space the bound 
$m_3^F\lsim$ TeV still persists. 

It has been noticed in \cite{Dorsner:2006fx}, that the constraint 
4) for $m_{(3,2)}^F$ can actually be evaded . 
In fact, there are solutions, in which 
$m_{(3,2)}^F\approx m_8^F/2$ that can be of order $M_{GUT}$. 
We have been however unable to find any solution with 
$M_{GUT}$ bigger than $10^{15}$ GeV, which makes them less 
realistic due to likely problems in large proton decay widths.

\begin{figure}[!ht]
  \centering
  \psfrag{x}{$\log_{10} \left(\frac{m_3}{\rm GeV}\right)$}
  \psfrag{y}{$\log_{10} \left(\frac{m_8}{\rm GeV}\right)$}
  \includegraphics[totalheight=5.5cm]{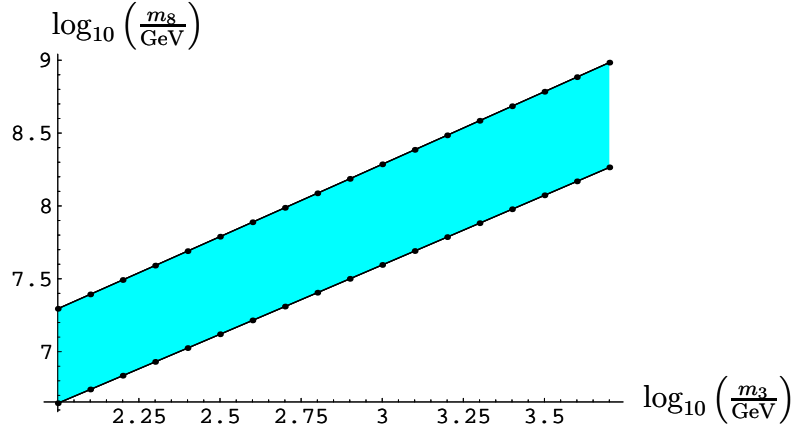}
\caption{\label{fig2} The region that gives 
unification at $M_{GUT}=10^{15.5}$ GeV.}
\end{figure}

Finally, one can ask, where must the octets be. Taking 
$M_{GUT}=10^{15.5}$ GeV one can find the possible region in 
$m_3-m_8$ plane, that leads 
to unification (different solutions for $m_8$ for the same $m_3$ 
correspond to different choices of $m_{(3,2)}^F$). This region is 
shown in Fig. \ref{fig2}.

\section{Phenomenological implications: testing seesaw at LHC?}

In the previous section we learned that the triplets are quite light, even 
likely to be found at LHC. How would they be identified?

The Yukawa couplings of the triplet and singlet fermion are 
parametrized by (we choose the basis in which the Dirac
Yukawa matrix between $e^c$ and $L$ is diagonal and real,
while $y_T^i$ are real)

\begin{eqnarray}
\label{lyukawa}
{\cal L}_Y&=&
-y_E^iH^\dagger e^c_iL_i
+y_T^iH^Ti\tau^2\tau^aT^aL_i
+y_S^iH^Ti\tau^2SL_i+h.c.\nonumber\\
&=&-\frac{v+h}{\sqrt{2}}\left[y_E^ie^c_ie_i
+y_T^i\left(\sqrt{2}T^+e_i+T^0\nu_i\right)
+y_S^iS\nu_i\right]+h.c.
\end{eqnarray}

\noindent
where $T^{\pm}$, $T^0$ are the three states from the fermionic 
triplet, while $S$ is the fermionic singlet. 
We have changed the cumbersome notation from the previous 
section (where it was necessary), since this whole section is 
devoted only to the fermionic triplet and singlet. 

The Majorana masses for the triplet
and singlet (with properly defined $T^k$ and $S$ the masses
$m_T$ and $m_S$ can be made real and positive) are

\begin{equation}
{\cal L}_{m}=-\frac{m_T}{2}\left(2T^+T^-+T^0T^0\right)
-\frac{m_S}{2}SS+h.c.
\end{equation}

To the leading order in the neutrino Dirac Yukawa couplings 
the following transformations define the physical states: 

\begin{eqnarray}
\nu_j&\to&\nu_j+\epsilon_T^jT^0+
\epsilon_S^jS\;,\\
T^0&\to&T^0-\epsilon_T^k\nu_k\;,\\
S&\to&S-\epsilon_S^k\nu_k\;,\\
e_j&\to&e_j+\sqrt{2}\epsilon_T^jT^-\;,\\
T^-&\to&T^--\sqrt{2}\epsilon_T^ke_k\;,\\
T^+&\to&T^+ , \; e^c \to e^c
\end{eqnarray}

\noindent
where 

\begin{equation}
\epsilon_X^i\equiv \frac{y_X^i v}{\sqrt{2} m_X}\;.
\end{equation}

In the above equation recall that $T^+$ is a different state 
from $T^-$, just like $e^c$ is a different state from $e$.

The light neutrino mass matrix is then given by 

\begin{equation}
\label{neutrinomass}
m^\nu_{ij}=-\frac{v^2}{2}\left(
\frac{y_T^iy_T^j}{m_T}+
\frac{y_S^iy_S^j}{m_S}\right)\;
\end{equation}

\noindent
in the basis in which the charged Yukawas and the couplings 
with $W$ are diagonal. 

\subsection{$T\to W(Z)+\;{\rm light \; lepton}$}

These are the predominant decay modes of the triplets, 
whose strength is dictated by the neutral Dirac Yukawa couplings. 

\begin{eqnarray} 
\label{t-z}
\Gamma(T^-\to Ze_k^-)&=&
\frac{m_T}{32\pi}\left|y_T^k\right|^2 
\left(1-\frac{m_Z^2}{m_T^2}\right)^2\left(1+2\frac{m_Z^2}{m_T^2}\right)\;,\\
\label{t-w}
\sum_k\Gamma(T^-\to W^-\nu_k)&=&
\frac{m_T}{16\pi}\left(\sum_k\left|y_T^k\right|^2\right)
\left(1-\frac{m_W^2}{m_T^2}\right)^2\left(1+2\frac{m_W^2}{m_T^2}\right)\;,\\
\label{t0w}
\Gamma(T^0\to W^+e_k^-)&=&
\Gamma(T^0\to W^-e_k^+)=\nonumber\\
&=&\frac{m_T}{32\pi}\left|y_T^k\right|^2
\left(1-\frac{m_W^2}{m_T^2}\right)^2\left(1+2\frac{m_W^2}{m_T^2}\right)\;,\\
\label{t0z}
\sum_k\Gamma(T^0\to Z\nu_k)&=&
\frac{m_T}{32\pi}\left(\sum_k\left|y_T^k\right|^2\right)
\left(1-\frac{m_Z^2}{m_T^2}\right)^2\left(1+2\frac{m_Z^2}{m_T^2}\right)\;,
\end{eqnarray}

\noindent
where we averaged over initial polarizations and summed over final ones.
From (\ref{t0w}) one sees that the decays of $T^0$, just as those of 
righthanded neutrinos, violate lepton number. In a machine such as LHC one 
would typically produce a pair $T^+T^0$ (or $T^-T^0$), whose decays 
then allow for interesting $\Delta L=2$ signatures of same sign 
dileptons and 4 jets. This fairly SM background free signature is characteristic 
of any theory with righthanded neutrinos as discussed in \cite{Keung:1983uu}. 
The main point here is that these triplets are really predicted to be 
light, unlike in the case of righthanded neutrinos. The detailed 
analyses of the LHC signatures including the production, the decays and 
the background is now in progress \cite{abdussalam}. 

The decay rates above are rather sensitive to the Yukawa couplings which 
on the other hand can vary a lot. First of all, they are not directly 
related to the neutrino properties, and they are of course rather flavour 
dependent. The dominant rate goes through the largest Yukawa 
coupling which has an approximate lower limit of $\approx 5\times 10^{-7}$ 
from the atmospheric neutrino oscillations. 
This translates into the following upper limit for the lifetime 
of the dominant two-body triplet decay, for say $m_T\approx 300$ GeV

\begin{equation}
\label{taulifetime}
\tau_T\lsim 10^{-1}\;{\rm mm}\;.
\end{equation}

Measuring the above decays means in some sense checking
the seesaw parameters. Let's see in more detail this correspondence.
The situation with the singlet and triplet making the light
neutrino massive through the seesaw mechanism is analogous to
the situation with two righthanded neutrinos (for a recent 
review of this situation see \cite{Guo:2006qa}). 
Thus we can use the known relations \cite{Ibarra:2003up} 
(in the case of hierarchical case, i.e. $m^\nu_1=0$)

\begin{eqnarray}
\label{t1}
\frac{vy_T^{i*}}{\sqrt{2}}&=&i\sqrt{m_T}\left(
U_{i2}\sqrt{m^\nu_2}\cos{z}\pm
U_{i3}\sqrt{m^\nu_3}\sin{z}\right)\;,\\
\label{s1}
\frac{vy_S^{i*}}{\sqrt{2}}&=&-i\sqrt{m_S}\left(
U_{i2}\sqrt{m^\nu_2}\sin{z}\mp
U_{i3}\sqrt{m^\nu_3}\cos{z}\right)\;,
\end{eqnarray}

\noindent
or (in the case of inverse hierarchy, i.e. $m^\nu_3=0$))

\begin{eqnarray}
\label{t3}
\frac{vy_T^{i*}}{\sqrt{2}}&=&i\sqrt{m_T}\left(
U_{i1}\sqrt{m^\nu_1}\cos{z}\pm
U_{i2}\sqrt{m^\nu_2}\sin{z}\right)\;,\\
\label{s3}
\frac{vy_S^{i*}}{\sqrt{2}}&=&-i\sqrt{m_S}\left(
U_{i1}\sqrt{m^\nu_1}\sin{z}\mp
U_{i2}\sqrt{m^\nu_2}\cos{z}\right)\;,
\end{eqnarray}

\noindent
valid in a different basis than used before, since here
$y_T^i$ are not necessarily real. To compare with the previous
results one needs just to compute the absolute value
$\left|y_T^i\right|$ in (\ref{t1}), (\ref{t3}).

In the formulae above $z$ is a complex number, while $U$ 
the PMNS matrix, that diagonalizes the neutrino mass matrix 
(\ref{neutrinomass}) (for the experimental values and limits 
see \cite{Strumia:2006db})

\begin{eqnarray}
m^\nu=U^*\pmatrix{
  m^\nu_1
&
  0
&
  0
\cr
  0
&
  m^\nu_2
&
  0
\cr
  0
&
  0
&
  m^\nu_3
\cr}
U^\dagger
\end{eqnarray}

Suppose we could measure from $T$ decays the Yukawa couplings $y_T^i$.
Then, in the above formulae we have the following unknowns: one
complex number $z$ and two CP phases, assuming
that the $1-3$ mixing will be measured soon (keep in mind that 
there is one CP phase less in the case of one massless neutrino). In general
it is not possible to give much constraints or to make some nontrivial
checks, since one has $3$ real measurements (the absolute values
of $y_T^i$), but $4$ parameters available to describe them. In some
special cases however the above relations simplify and some nontrivial
constraints appear. 

As an example consider the inverse hierarchical case with a vanishing 
$\theta_{13}$. One gets 

\begin{equation}
\frac{\Gamma(\tau)}{\Gamma(\mu)}= \tan^2{\theta_{atm}}
\end{equation}

\noindent
independent on the phases. This can serve as a direct test of the theory 
if the inverse hierarchy and a small enough $\theta_{13}$ are to be 
established in the future. 

Another interesting case is $|Im(z)|\gg 1$, which is equivalent to the 
large Dirac Yukawa limit. Here the complex parameter $z$ disappears from 
the branching ratios, which then depend only on the in principle 
measurable parameters of the PMNS mixing matrix. 

\subsection{$T^\pm\to T^0$ decays}

For a nonvanishing and positive mass split 
$\Delta m_T\equiv m_{T^+}-m_{T^0}$ the charged triplet 
fermion can decay into a neutral one and an (off-shell) 
$W$. 

One gets for $\Delta m_T$ at the one-loop level 

\begin{equation}
\label{dmt}
\Delta m_T=\frac{\alpha_2}{2\pi}\frac{m_W^2}{m_T}
\left[f\left(\frac{m_T}{m_Z}\right)-f\left(\frac{m_T}{m_W}\right)\right]\;,
\end{equation}

\noindent
where

\begin{equation}
f\left(y\right)=\frac{1}{4y^2}\log{y^2}-
\left(1+\frac{1}{2y^2}\right)
\sqrt{4y^2-1}\arctan{\sqrt{4y^2-1}}\;,
\end{equation}

\noindent
which gives $\Delta m_T\approx 160$ MeV 
with $10\%$ accuracy in the whole range $m_Z\le m_T\le\infty$. 
Notice that there is also a possible direct tree-level contribution 
from (\ref{lf}) through a non-vanishing vev of the bosonic triplet 

\begin{equation}
\Delta m_T^{tree}\approx y\left<T_B\right>\;.
\end{equation}

However, $\left<T_B\right>\lsim 1$ GeV for the $W$ and $Z$ masses
and $y\lsim 10^{-2}$ since suppressed by $M_{GUT}/\Lambda$, 
so $\Delta m_T^{tree}\lsim 10$ MeV, a negligible addition. 

The fastest decay mode through the above mass difference is 
clearly $T^\pm\to T^0\pi^\pm$, 
estimated to be ${\cal O}(10^{-10})$ sec \cite{Ibe:2006de}, 
negligible in comparison with the $W^\pm\nu$ or $Z l^\pm$ decay 
channels considered in the previous subsection.

In short, the triplet mass difference can be safely ignored. 

\subsection{$T\to H+\;{\rm light \; lepton}$}

If the fermionic triplet is heavier than the SM Higgs, it 
can decay unsuppressed also to the Higgs and a light lepton. 
The decay widths can be calculated from (\ref{lyukawa}) to give

\begin{eqnarray}
\label{t-h}
\Gamma\left(T^-\to h e_k^-\right)&=&\frac{m_T}{32\pi}\left|y_T^k\right|^2
\left(1-\frac{ m_h^2  }{ m_T^2  }\right)^2\;,\\
\label{t0h}
\sum_k\Gamma\left(T^0\to h \nu_k\right)&=&\frac{m_T}{32\pi}
\left(\sum_k\left|y_T^k\right|^2\right)
\left(1-\frac{ m_h^2  }{ m_T^2  }\right)^2\;.
\end{eqnarray}

These results can now explain the apparent ``puzzle'' 
from the results (\ref{t-z})-(\ref{t0z}). In fact, these decays 
come out to be nonzero also in the SU(2) preserving limit ($v\to 0$). 
However, in this limit there is no mixing between the triplets 
and the light leptons, so apparently no decays. The results 
(\ref{t-h})-(\ref{t0h}) explains the discrepancy: in this limit 
there are four degrees of freedom from the Higgs doublet and 
the final states in (\ref{t-z})-(\ref{t0z}) should be interpreted 
as $Z$ being the imaginary partner of the standard Higgs, and 
$W$ being the complex partner in the doublet (the upper component). 
It is easy to check that the exact SU(2) gauge symmetry connects 
the results (\ref{t-z})-(\ref{t0z}) with (\ref{t-h})-(\ref{t0h}) 
in the limiting case $v,m_h\to 0$.

\section{Cosmological implications}

\subsection{Dark matter}

As usual, in order to have a viable dark matter candidate,
it must be stable for at least the age of the universe, which 
can be translated into an extremely small decay width:

\begin{equation}
\Gamma\lsim 10^{-42}\;{\rm GeV}\;.
\end{equation}  

Let us systematically consider various possible candidates:

1) the fermionic neutral triplet $T^0$: obviously this cannot work,
see (\ref{taulifetime});

2) next consider the bosonic triplet from $24_H$, with a mass
of at least $m_T^B\ge 100$ GeV from collider constraints. Now, the 
following operator

\begin{equation}
\bar 5_F\frac{24_H}{\Lambda}10_F5_H^\dagger
\end{equation}  

\noindent
is needed to correct the $b-\tau$ unification, as discussed in the previous
 section. So the bosonic triplet
can decay into a fermion antifermion pair with a decay width of

\begin{equation}
\Gamma\approx \left(\frac{m_W}{\Lambda}\right)^2m_T^B
\end{equation}

\noindent
much too fast ($\approx 10^{-32}$ GeV) even in the unrealistic case
of $\Lambda = M_{Pl}$.

3) what about the bosonic singlet in $24_H$?
This singlet is nothing else than the field that breaks
SU(5), so the validity of the whole approach 
requires  its mass to be larger than the electroweak scale. Then
everything of the previous case applies also here, and thus
the same negative conclusion.

4) finally the fermionic singlet $S$ in $24_F$. At first sight,
this could not work for the same reason as for the fermionic
triplet $T$. However, here there are no such tight constraints
on its mass from colliders, so in principle it could weigh 
around keV. In this case its decay rate gets suppressed 
strongly by the $W$ propagator, giving

\begin{equation}
\Gamma\approx\left(y_S^iG_F\right)^2m_S^5\;,
\end{equation}

\noindent
which is slow enough in the keV region as soon as $y_S^i\lsim 0.1$,
needed anyway for unitarity. Unfortunately, in order to have the right 
amount of dark matter, such a small mass cannot give a sizeable 
contribution to the neutrino masses (see \cite{Asaka:2005an} for details).

\subsection{Leptogenesis}

This issue has been discussed at length in a fairly 
complete paper devoted to the phenomenology and cosmology of the 
type I seesaw mechanism with only two righthanded neutrinos 
\cite{Guo:2006qa}. 
The assumption (two $\nu_R$) in that case 
is a prediction of our model ($T$ and $S$). The bottom line 
now is that, as opposed to the generic situation with three 
righthanded neutrinos 
\footnote{for a way out of the well known Davidson-Ibarra limit 
\cite{Davidson:2002qv}, see \cite{Hambye:2003rt,Raidal:2004vt}}, 
there is a true physical lower limit 
on the scale of leptogenesis of the order $10^{10}$ GeV. 
This can be seen from a straightforward derivation of the 
following expression for the maximal CP asymmetry (assuming 
a hierarchy $m_T\ll m_S$)

\begin{equation}
\epsilon_{MAX}\approx \frac{1}{16\pi}\frac{m_T\left(m^\nu_3-m^\nu_2\right)}
{v^2}\;.
\end{equation}

It is evident that for $m_T\approx$ TeV the CP asymmetry is hopelessly 
small. The only possibility for leptogenesis in this theory is 
the resonant \cite{Flanz:1996fb,Pilaftsis:1997jf,Pilaftsis:2003gt} one. 
It is not difficult to show that 
one can get realistic value for the baryon asymmetry. Following 
\cite{Hambye:2003rt} the triplet asymmetry (and similarly the singlet 
one) can be rewritten as 

\begin{equation}
\epsilon_T=-\frac{Im
\left(\overrightarrow{y_T}\overrightarrow{y_S}^{*}\right)^2}
{2\left|\overrightarrow{y_T}\right|^2
\left|\overrightarrow{y_S}\right|^2}\times
\frac{2\left(m_S^2-m_T^2\right)m_T\Gamma_S}
{\left(m_S^2-m_T^2\right)^2+m_T^2\Gamma_S^2}\;.
\end{equation}

\noindent
where $\Gamma_S$ is the total decay width of the singlet fermion. 
The last term in the above product can take its maximal value $1$ 
for properly chosen singlet mass, i.e. for $m_S^2=m_T^2+m_T\Gamma_S$. 
The first term can be rewritten using (\ref{t1})-(\ref{s3}) and 
depends on the value of the unknown complex parameter $z$ and the 
parameters (masses, mixing angles and phases) of the light neutrino sector. 
This term can be numerically even of order one, showing that a very 
large asymmetry can be obtained. It has to be stressed however that 
successfull resonant leptogenesis does not allow all values of $z$. 
For example, if $Re(z)$ or $Im(z)$ is zero, the final result vanishes. 
It is interesting that the asymmetry generically decreases with large 
$Im(z)$ values, restricting the allowed region. A more detailed 
description with the calculation of the efficiency factor and the 
inclusion of flavour effects \cite{Pascoli:2006ie,Pascoli:2006ci} 
is beyond the scope of this paper and is in progress \cite{fcnclepto}. 
Preliminary estimates seem to show that these restrictions are more 
restrictive than the ones coming from flavour changing neutral processes.

The requirement of successfull leptogenesis thus puts various constraints 
on the yet unknown model parameters. Although the degeneracy between the 
singlet and the triplet mass is not of direct meaning for LHC searches 
(it may be relevant however for processes under study \cite{fcnclepto} 
like $\mu\to e\gamma$), because the singlets will not be easily produced, 
the constraints on the Yukawas could be tested measuring the different 
branching ratios in triplet decays. 

\section{Summary and outlook}

We have recently \cite{Bajc:2006ia} constructed what can be 
considered a possible minimal realistic SU(5) grand unified 
theory. Instead of changing the Higgs sector as conventionally 
done,  we have simply added to the minimal model an adjoint fermion 
representation which gives a hybrid seesaw scenario, a mixture of 
type I and type III. The fact that the theory is in accord with 
experiment may not be so surprirising; after all one has enlarged its
particle sector. What is remarkable is a prediction 
of a light fermionic SU(2) triplet, with a mass below TeV, and for 
a large portion of the parameter space in the LHC reach of below 
500 GeV. One has a badly needed predictive theory of seesaw 
mechanism that can be tested at the collider energies. 
 
Whereas it is always possible to imagine a low energy seesaw, 
predictive theories such as GUTs normally prefer large seesaw 
scale, close to $M_{GUT}$.  Even when you assume this scale to be low 
as often done in the type I case, the production of righthanded 
neutrinos is suppressed by the small (compared to the gauge couplings) 
Yukawa couplings (for a recent work see \cite{Kersten:2007vk}). 
The type II could of course be tested for a low 
scale, but again that is not what comes out. In a way, type III seesaw, 
up to now almost not studied at all, may provide a unique possibility 
of seesaw tested at LHC.

In this longer version of our letter we have carefully studied 
some phenomenological and cosmological issues in this theory. 
We have performed a complete two-loop analysis of unification 
constraints which confirms the lightness of the triplets, and 
at the same time predicts $M_{GUT} < 10^{16} $ GeV, implying the 
proton lifetime below $10^{36}$ years, possibly observable in the 
next generation of proton decay experiments. We have discussed the decay
of the triplets into the charged leptons and neutrinos and shown 
how they probe directly neutrino Dirac Yukawa couplings. 
These Dirac Yukawas could be
quite large since small neutrino masses can involve cancellations 
and in this case lead to possibly observable lepton flavor violating 
processes. This is a rather interesting topic and deserves a careful 
investigation in a separate note. 
We also confirmed here an expected 
result that  only resonant leptogenesis can work due to the low mass 
of the fermionic triplet. This would also make a fermionic singlet 
light; but as in the case of pure type I seesaw it is not  of direct 
phenomenological interest. Finally, we also showed that the theory 
lacks a dark matter candidate. Of course, one could always add a 
singlet particle and account for the dark matter if necessary, 
since for a given mass one can choose an appropriate coupling.

It is interesting to compare this model to the supersymmetric 
SU(5) theory, or the supersymmetric extension of the standard model. 
After all, the weak triplet fermions correspond to winos, while the 
color octet fermions in $24_H$ correspond to gluinos. Now, it is 
well known that the sfermions do not enter into the renormalization 
group constraints, at least not at the one-loop level; they can be as 
heavy as we wish. This,  split supersymmetry program:  
light winos, gluinos and higgsinos still allows for the unification 
of gauge couplings, as much as in the case of low energy supersymmetry 
\cite{Dimopoulos:1981yj,Ibanez:1981yh,Einhorn:1981sx,Marciano:1981un}. 
Our work shows that the situation can be more complex if one is
willing to split superymmetry: one can have higgsinos completely 
decoupled, and the gluinos in the intermediate region. But then, 
by interpolation, there is clearly a continuum of solutions with
higgsino anywhere from the weak to the Planck scale, and gluinos from the 
weak to some intermediate scale. The work on this in progress and will 
be reported elsewhere \cite{doublesplitsusy}. It is interesting though 
that LHC may see only winos and nothing else if supersymmetry is to be 
split, similar as in the theory discussed here. The important difference 
in our case is the fact that the fermionic triplet, an analogue of 
winos, is directly related to neutrino masses and mixings. It 
should be viewed as a possible alternative to low energy 
supersymmetry; instead of not so well defined principle of 
naturalness, it has direct physical and phenomenological motivation. 

\section*{Acknowledgements}

The work of G.S. was supported in part by the European Commission under 
the RTN contract MRTN-CT-2004-503369; the work of B.B. and M.N. was 
supported by the Slovenian Research Agency. B.B. and M.N. 
thank ICTP for hospitality during the course of this work. 
We thank Paolo Creminelli, Ilja Dor\v sner, Tsedenbaljir Enkhbat, 
Alejandra Melfo, Fabrizio Nesti and Francesco Vissani for discussion, 
and Marco Cirelli, Michele Frigerio 
and Alessandro Strumia for pointing out a sign error in eq. (\ref{dmt}) 
of the first version of the manuscript.

\end{document}